\RequirePackage{lineno}
\documentclass[aps,prl,preprint,tightenlines,superscriptaddress,showpacs,byrevtex,nofootinbib]{revtex4-1}
\usepackage{graphicx} 
\usepackage{dcolumn}  
\usepackage{amssymb}
\usepackage{xcolor}
\graphicspath{{ps}}

\begin{document}


\preprint{\vbox{ \hbox{   }
                 \hbox{BELLE-CONF-1805}
}}

\title{ \quad\\[0.5cm] Measurement of the $D^{\ast-}$ polarization in
the decay $B^0 \to D^{\ast -}\tau^+\nu_{\tau}$}

\noaffiliation
\affiliation{University of the Basque Country UPV/EHU, 48080 Bilbao}
\affiliation{Beihang University, Beijing 100191}
\affiliation{University of Bonn, 53115 Bonn}
\affiliation{Brookhaven National Laboratory, Upton, New York 11973}
\affiliation{Budker Institute of Nuclear Physics SB RAS, Novosibirsk 630090}
\affiliation{Faculty of Mathematics and Physics, Charles University, 121 16 Prague}
\affiliation{Chiba University, Chiba 263-8522}
\affiliation{Chonnam National University, Kwangju 660-701}
\affiliation{University of Cincinnati, Cincinnati, Ohio 45221}
\affiliation{Deutsches Elektronen--Synchrotron, 22607 Hamburg}
\affiliation{Duke University, Durham, North Carolina 27708}
\affiliation{University of Florida, Gainesville, Florida 32611}
\affiliation{Department of Physics, Fu Jen Catholic University, Taipei 24205}
\affiliation{Key Laboratory of Nuclear Physics and Ion-beam Application (MOE) and Institute of Modern Physics, Fudan University, Shanghai 200443}
\affiliation{Justus-Liebig-Universit\"at Gie\ss{}en, 35392 Gie\ss{}en}
\affiliation{Gifu University, Gifu 501-1193}
\affiliation{II. Physikalisches Institut, Georg-August-Universit\"at G\"ottingen, 37073 G\"ottingen}
\affiliation{SOKENDAI (The Graduate University for Advanced Studies), Hayama 240-0193}
\affiliation{Gyeongsang National University, Chinju 660-701}
\affiliation{Hanyang University, Seoul 133-791}
\affiliation{University of Hawaii, Honolulu, Hawaii 96822}
\affiliation{High Energy Accelerator Research Organization (KEK), Tsukuba 305-0801}
\affiliation{J-PARC Branch, KEK Theory Center, High Energy Accelerator Research Organization (KEK), Tsukuba 305-0801}
\affiliation{Forschungszentrum J\"{u}lich, 52425 J\"{u}lich}
\affiliation{Hiroshima Institute of Technology, Hiroshima 731-5193}
\affiliation{IKERBASQUE, Basque Foundation for Science, 48013 Bilbao}
\affiliation{University of Illinois at Urbana-Champaign, Urbana, Illinois 61801}
\affiliation{Indian Institute of Science Education and Research Mohali, SAS Nagar, 140306}
\affiliation{Indian Institute of Technology Bhubaneswar, Satya Nagar 751007}
\affiliation{Indian Institute of Technology Guwahati, Assam 781039}
\affiliation{Indian Institute of Technology Hyderabad, Telangana 502285}
\affiliation{Indian Institute of Technology Madras, Chennai 600036}
\affiliation{Indiana University, Bloomington, Indiana 47408}
\affiliation{Institute of High Energy Physics, Chinese Academy of Sciences, Beijing 100049}
\affiliation{Institute of High Energy Physics, Vienna 1050}
\affiliation{Institute for High Energy Physics, Protvino 142281}
\affiliation{Institute of Mathematical Sciences, Chennai 600113}
\affiliation{INFN - Sezione di Napoli, 80126 Napoli}
\affiliation{INFN - Sezione di Torino, 10125 Torino}
\affiliation{Advanced Science Research Center, Japan Atomic Energy Agency, Naka 319-1195}
\affiliation{J. Stefan Institute, 1000 Ljubljana}
\affiliation{Kanagawa University, Yokohama 221-8686}
\affiliation{Institut f\"ur Experimentelle Teilchenphysik, Karlsruher Institut f\"ur Technologie, 76131 Karlsruhe}
\affiliation{Kavli Institute for the Physics and Mathematics of the Universe (WPI), University of Tokyo, Kashiwa 277-8583}
\affiliation{Kennesaw State University, Kennesaw, Georgia 30144}
\affiliation{King Abdulaziz City for Science and Technology, Riyadh 11442}
\affiliation{Department of Physics, Faculty of Science, King Abdulaziz University, Jeddah 21589}
\affiliation{Kitasato University, Sagamihara 252-0373}
\affiliation{Korea Institute of Science and Technology Information, Daejeon 305-806}
\affiliation{Korea University, Seoul 136-713}
\affiliation{Kyoto University, Kyoto 606-8502}
\affiliation{Kyungpook National University, Daegu 702-701}
\affiliation{LAL, Univ. Paris-Sud, CNRS/IN2P3, Universit\'{e} Paris-Saclay, Orsay}
\affiliation{\'Ecole Polytechnique F\'ed\'erale de Lausanne (EPFL), Lausanne 1015}
\affiliation{P.N. Lebedev Physical Institute of the Russian Academy of Sciences, Moscow 119991}
\affiliation{Liaoning Normal University, Dalian 116029}
\affiliation{Faculty of Mathematics and Physics, University of Ljubljana, 1000 Ljubljana}
\affiliation{Ludwig Maximilians University, 80539 Munich}
\affiliation{Luther College, Decorah, Iowa 52101}
\affiliation{Malaviya National Institute of Technology Jaipur, Jaipur 302017}
\affiliation{University of Malaya, 50603 Kuala Lumpur}
\affiliation{University of Maribor, 2000 Maribor}
\affiliation{Max-Planck-Institut f\"ur Physik, 80805 M\"unchen}
\affiliation{School of Physics, University of Melbourne, Victoria 3010}
\affiliation{University of Mississippi, University, Mississippi 38677}
\affiliation{University of Miyazaki, Miyazaki 889-2192}
\affiliation{Moscow Physical Engineering Institute, Moscow 115409}
\affiliation{Moscow Institute of Physics and Technology, Moscow Region 141700}
\affiliation{Graduate School of Science, Nagoya University, Nagoya 464-8602}
\affiliation{Kobayashi-Maskawa Institute, Nagoya University, Nagoya 464-8602}
\affiliation{Universit\`{a} di Napoli Federico II, 80055 Napoli}
\affiliation{Nara University of Education, Nara 630-8528}
\affiliation{Nara Women's University, Nara 630-8506}
\affiliation{National Central University, Chung-li 32054}
\affiliation{National United University, Miao Li 36003}
\affiliation{Department of Physics, National Taiwan University, Taipei 10617}
\affiliation{H. Niewodniczanski Institute of Nuclear Physics, Krakow 31-342}
\affiliation{Nippon Dental University, Niigata 951-8580}
\affiliation{Niigata University, Niigata 950-2181}
\affiliation{University of Nova Gorica, 5000 Nova Gorica}
\affiliation{Novosibirsk State University, Novosibirsk 630090}
\affiliation{Osaka City University, Osaka 558-8585}
\affiliation{Osaka University, Osaka 565-0871}
\affiliation{Pacific Northwest National Laboratory, Richland, Washington 99352}
\affiliation{Panjab University, Chandigarh 160014}
\affiliation{Peking University, Beijing 100871}
\affiliation{University of Pittsburgh, Pittsburgh, Pennsylvania 15260}
\affiliation{Punjab Agricultural University, Ludhiana 141004}
\affiliation{Research Center for Electron Photon Science, Tohoku University, Sendai 980-8578}
\affiliation{Research Center for Nuclear Physics, Osaka University, Osaka 567-0047}
\affiliation{Theoretical Research Division, Nishina Center, RIKEN, Saitama 351-0198}
\affiliation{RIKEN BNL Research Center, Upton, New York 11973}
\affiliation{Saga University, Saga 840-8502}
\affiliation{University of Science and Technology of China, Hefei 230026}
\affiliation{Seoul National University, Seoul 151-742}
\affiliation{Shinshu University, Nagano 390-8621}
\affiliation{Showa Pharmaceutical University, Tokyo 194-8543}
\affiliation{Soongsil University, Seoul 156-743}
\affiliation{University of South Carolina, Columbia, South Carolina 29208}
\affiliation{Stefan Meyer Institute for Subatomic Physics, Vienna 1090}
\affiliation{Sungkyunkwan University, Suwon 440-746}
\affiliation{School of Physics, University of Sydney, New South Wales 2006}
\affiliation{Department of Physics, Faculty of Science, University of Tabuk, Tabuk 71451}
\affiliation{Tata Institute of Fundamental Research, Mumbai 400005}
\affiliation{Excellence Cluster Universe, Technische Universit\"at M\"unchen, 85748 Garching}
\affiliation{Department of Physics, Technische Universit\"at M\"unchen, 85748 Garching}
\affiliation{Toho University, Funabashi 274-8510}
\affiliation{Tohoku Gakuin University, Tagajo 985-8537}
\affiliation{Department of Physics, Tohoku University, Sendai 980-8578}
\affiliation{Earthquake Research Institute, University of Tokyo, Tokyo 113-0032}
\affiliation{Department of Physics, University of Tokyo, Tokyo 113-0033}
\affiliation{Tokyo Institute of Technology, Tokyo 152-8550}
\affiliation{Tokyo Metropolitan University, Tokyo 192-0397}
\affiliation{Tokyo University of Agriculture and Technology, Tokyo 184-8588}
\affiliation{Utkal University, Bhubaneswar 751004}
\affiliation{Virginia Polytechnic Institute and State University, Blacksburg, Virginia 24061}
\affiliation{Wayne State University, Detroit, Michigan 48202}
\affiliation{Yamagata University, Yamagata 990-8560}
\affiliation{Yonsei University, Seoul 120-749}
  \author{A.~Abdesselam}\affiliation{Department of Physics, Faculty of Science, University of Tabuk, Tabuk 71451} 
  \author{I.~Adachi}\affiliation{High Energy Accelerator Research Organization (KEK), Tsukuba 305-0801}\affiliation{SOKENDAI (The Graduate University for Advanced Studies), Hayama 240-0193} 
  \author{K.~Adamczyk}\affiliation{H. Niewodniczanski Institute of Nuclear Physics, Krakow 31-342} 
  \author{J.~K.~Ahn}\affiliation{Korea University, Seoul 136-713} 
  \author{H.~Aihara}\affiliation{Department of Physics, University of Tokyo, Tokyo 113-0033} 
  \author{S.~Al~Said}\affiliation{Department of Physics, Faculty of Science, University of Tabuk, Tabuk 71451}\affiliation{Department of Physics, Faculty of Science, King Abdulaziz University, Jeddah 21589} 
  \author{K.~Arinstein}\affiliation{Budker Institute of Nuclear Physics SB RAS, Novosibirsk 630090}\affiliation{Novosibirsk State University, Novosibirsk 630090} 
  \author{Y.~Arita}\affiliation{Graduate School of Science, Nagoya University, Nagoya 464-8602} 
  \author{D.~M.~Asner}\affiliation{Brookhaven National Laboratory, Upton, New York 11973} 
  \author{H.~Atmacan}\affiliation{University of South Carolina, Columbia, South Carolina 29208} 
% ...
\collaboration{The Belle Collaboration}

\begin{abstract}
We report the first measurement of the $D^{\ast -}$ meson polarization in
the decay $B^0 \to D^{*-} \tau^+\nu_{\tau}$ using the full data sample
of 772$\times 10^6$ $B\bar{B}$ pairs recorded with the Belle detector
at the KEKB
electron-positron collider.  Our result, $F_L^{D^\ast} = 0.60 \pm 0.08
({\rm stat}) \pm 0.04 ({\rm sys})$, where $F_L^{D^\ast}$ denotes the 
$D^{\ast-}$
meson
longitudinal polarization fraction, agrees within about $1.7$ standard 
deviations of the standard model prediction.
\end{abstract}


\maketitle


{\renewcommand{\thefootnote}{\fnsymbol{footnote}}}
\setcounter{footnote}{0}

%
\section{Introduction}

Decays of $B$ mesons to final states containing $\tau$ leptons provide an
important test-bed for the standard model (SM) and its extensions. Of
special interest are theoretically well-controlled semitauonic decays
$B\to \bar{D}^{(\ast)}\tau^+\nu_{\tau}$ \cite{CG}, where
new physics (NP) may contribute at tree level.  Complementary 
sensitivities  of
the decays $B\to \bar{D}\tau^+\nu_{\tau}$ and $B\to
\bar{D}^{\ast}\tau^+\nu_{\tau}$ to various SM extensions, and  the rich
spectrum of kinematical observables accessible in the three-body final
states,
enable  comprehensive studies of the underlying dynamics in $\bar{b}\to
\bar{c}\tau^+\nu_{\tau}$ transitions \cite{Fajfer_1, bstd}. This 
potential is
still  far from being fully explored, primarily due to the inherent
measurement challenges associated with multiple neutrino final states.

The decays $B\to \bar{D}^{(\ast)}\tau^+\nu_{\tau}$  have been 
studied
experimentally by Belle \cite{AM, AB, MH, YS, SH}, BaBar
\cite{BaBar1,BaBar2}, and LHCb \cite{LHCb1,LHCb2}. 
So far, the
experiments measured the branching fractions
$\mathcal{B}(B\to
\bar{D}^{(\ast)}\tau^+\nu_{\tau})$, or the ratios
$R(D^{(\ast)})=\mathcal{B}(B\to
\bar{D}^{(\ast)}\tau^+\nu_{\tau})/\mathcal{B}(B\to
\bar{D}^{(\ast)}\ell^+\nu_{\ell})$,  ($\ell=e, \mu$),
distributions of several kinematic variables,
and recently the longitudinal tau
polarization, $P_{\tau}^{D^{\ast}}$, in the $D^{\ast}$ mode \cite{SH}. While the
results on differential decay rates and $P_{\tau}^{D^{\ast}}$ are still
statistically limited, the experimental values of  $R(D^{(\ast)})$
already challenge the SM and some of its extensions. The current world averages of $R(D)=0.407\pm 0.039\pm 0.024$ 
and
$R(D^{\ast})=0.306\pm 0.013\pm 0.007$ \cite{HFAG} exceed the SM
predictions
$R(D)=0.299\pm 0.003$ \cite{RD}, and $R(D^{\ast})=0.257\pm 0.003$
\cite{RDst}
by 2.3 and 3.0 standard deviations ($\sigma$), respectively, and the
combined results on $R(D^{(\ast)})$ deviate from the SM by about
3.8 $\sigma$.
Interestingly, it is also difficult to accommodate the observed 
branching
fractions within the two Higgs doublet models \cite{celis, HDM2},
mainly due to the relatively large excess in the $B\to
\bar{D}^{\ast}\tau^+\nu_{\tau}$ mode, which is expected to be less
sensitive to
the charged Higgs contributions than the $B\to \bar{D}\tau^+\nu_{\tau}$
channel.   Further studies of kinematic distributions and angular
observables in semitauonic $B$ decays may provide new clues to unravel the
$R(D^{(\ast)})$ puzzle. An interesting observable, not explored so far
experimentally, is the $D^*$ polarization. In the SM,   
the fraction of $D^{\ast}$ longitudinal polarization, 
$F_L^{D^{\ast}}$, is expected to be around $0.45$
\cite{bstd, ivanov, alok, huang, Srimoy}, 
and the most recent
predictions are 
$0.441 \pm 0.006$ \cite{huang},
and $0.457 \pm 0.010$ \cite{Srimoy}.
The value of $F_L^{D^{\ast}}$ can be significantly modified in the presence of NP contributions
\cite{bstd, alok, huang, Srimoy, ivanov2}; in particular, the scalar 
and tensor operators
may enhance and decrease $F_L^{D^{\ast}}$, respectively. In this paper, we present the first measurement of the $D^*$  polarization  
in
the $B^0 \to D^{\ast -} \tau^+\nu_{\tau}$  decay. We extract $F_L^{D^{\ast}}$ from the angular distribution in
$D^{\ast -}\to \bar{ D}^0 \pi^-$ decay:
\begin{equation}
\label{eq:fl}
\frac{1}{\Gamma}\frac{d\Gamma}{d\cos\theta_{\rm hel}}
=\frac{3}{4}(2F_L^{D^{\ast}}\cos^2\theta_{\rm hel} + (1-F_L^{D^{\ast}})\sin^2\theta_{\rm 
hel}),
\end{equation}
where $\theta_{\rm hel}$ is the angle between $\bar{D^0}$ and 
the direction opposite to $B^0$ in
the $D^{\ast -}$ rest frame.

\section{\label{sec:detect}Detector and data samples}

This analysis is based on the full $\Upsilon(4S)$ data sample containing
$772\times 10^6\bar{B} B$ pairs recorded with the Belle detector
at the asymmetric-beam-energy $e^+e^-$ collider KEKB
\cite{KEKB}. The Belle detector, described in detail elsewhere
\cite{BelleDet}, is a
large-solid-angle magnetic spectrometer that consists of a silicon vertex
detector (SVD), a 50-layer central drift chamber (CDC), an array of
aerogel threshold Cherenkov counters (ACC), a barrel-like arrangement of
time-of-flight scintillation counters (TOF), and an electromagnetic
calorimeter (ECL) comprised of CsI(Tl) crystals located inside a
superconducting solenoid coil that provides a 1.5 T magnetic field.  An
iron flux-return located outside of the coil is instrumented to detect
$K^0_L$ mesons and to identify muons (KLM). Two inner detector
configurations were used. A 2.0 cm radius beampipe and a 3-layer SVD was
used for the first sample of $152\times 10^6$ $B\bar{B}$ pairs, while a
1.5 cm radius beampipe, a 4-layer SVD and a small-cell inner drift chamber
were used to record the remaining $620\times 10^6$ $B\bar{B}$ pairs
\cite{ZN}.
The analysis procedure is established using Monte Carlo (MC) samples.
Particle decays are modeled by the EvtGen package \cite{evtgen}, and
followed by detector simulation performed with GEANT3 \cite{geant}.
Radiative effects are modeled by PHOTOS \cite{photos}. Two large
samples ($100\times 10^6$ events  each) of $B^0 \to
D^{*-}\tau^+\nu_{\tau}$ decays are simulated within the SM using hadronic
form factors based on the Isgur-Scora-Grinstein-Wise (ISGW) model \cite{isgw2}
and on heavy quark effective theory (HQET) \cite{bstd}, respectively. 
To model the
background, we use MC
samples
of continuum $q\bar{q}$ ($q = u, d, s, c$), and
inclusive $B\bar{B}$ decays.
The sizes of these samples are, respectively, six and ten times
that of the collision data. Additionally, we use a sample of semileptonic $B$ decays to
orbitally-excited charmed mesons
$B \to \bar{D}^{\ast\ast}\ell^+\nu_{\ell}$ ($\bar{D}^{\ast\ast}$ stands
for
$\bar{D}_1$, $\bar{D}^{\ast}_2$, $\bar{D}_1'$, and $\bar{D}^{\ast}_0$)
generated with the ISGW
model and decay kinematics corrected to match
the Leibovich-Ligeti-Stewart-Wise model \cite{LLSW}, that exceeds six times 
the 
data sample.
\section{\label{sec:signal}Signal reconstruction}

The analysis adopts the approach of Refs.~\cite{AM, AB};  signal decays
($B_{\rm sig}$) are reconstructed first, and the accompanying $B$ meson
($B_{\rm tag}$) is
reconstructed inclusively from all the particles that remain after
selecting the $B_{\rm sig}$ candidate.
We use the following
secondary $B_{\rm sig}$ decays: $\tau^+ \to \ell ^+ \nu_{\ell}
\bar{\nu}_{\tau}$,
$\pi ^+ \bar{\nu}_{\tau}$,
$D^{*-}\to \bar{D}^0\pi^-$, $\bar{D^0}\to K^+\pi^-$,
$K^+\pi^-\pi^0$, $K^+\pi^+\pi^-\pi^-$ (denoted hereinafter as $K\pi$,
$K\pi\pi^0$ and $K3\pi$, respectively).
Primary charged tracks are required to have impact parameters consistent
with an origin at the interaction point (IP), and to have momenta 
in the laboratory frame
above 50 MeV \cite{units}. 
$K_S^0$ mesons
are reconstructed using pairs of charged tracks (treated as pions) satisfying
the invariant mass requirement
482 ${\rm MeV} <
M_{\pi^+\pi^-}< 514$ MeV  with a vertex displacement from the IP
consistent with the reconstructed momentum vector. Muons, electrons,
charged pions, kaons and protons are identified using information from the particle identification subsystems. The momenta of particles
identified as electrons are corrected for bremsstrahlung by adding photons
within a 50 mrad cone along the lepton trajectory.
The  $\pi^0$ candidates are reconstructed from photon pairs having 118 MeV
$< M_{\gamma\gamma} <$150 MeV. For candidates that share a
common
$\gamma$, we select the one with the smallest $\chi^2$ value resulting
from a $\pi^0$ mass-constrained fit. To reduce the combinatorial
background, we require that the photons from the $\pi^0$ have energies
greater than 50 MeV in the barrel part of the ECL
and greater than 100 MeV in the end-caps.
Photons that are not associated with a $\pi^0$ are accepted if their
energy exceeds a polar-angle dependent threshold ranging from 100 MeV to
200 MeV.

The $\bar{D}^0$ candidates, formed in the above specified channels, are
required to have masses in the range
$-25(-30)$ ${\rm MeV} < M_{\bar{D}^0}-m_{\bar{D}^0} < 25$ MeV for
$\bar{D}^0\to
K\pi, K3\pi (K\pi\pi^0)$ around the nominal $\bar{D}^0$ mass,
$m_{\bar{D}^0}$ \cite{pdg},  corresponding to a window of
approximately $\pm 4.5(\pm 2.5)\sigma$.  The $D^{\ast -}$
candidates
are reconstructed from $\bar{D}^0\pi^-$ pairs; we require that the
mass difference $\Delta M_{D^{\ast}} = M_{D^{\ast -}}- M_{\bar{D}^0}$ lie in the
window  $\pm 2.5$ MeV ($\pm 3\sigma$) around the nominal value of 145.43 MeV \cite{pdg}.

Signal candidates are selected by combining a $D^{\ast -}$ meson with
an oppositely charged electron, muon or pion. In the sub-channels with the
$\tau^+ \to \pi^+\bar{\nu}_{\tau}$ decay, the large combinatorial
background is suppressed by requiring the pion energy to be 
more than 0.5 GeV.
Particles that are not assigned to $B_{\rm sig}$ are used to 
reconstruct
the $B_{\rm tag}$ decay.
The consistency of a $B_{\rm tag}$ candidate with a
$B$-meson decay is checked using the beam-energy constrained mass and the
energy difference variables in the $\Upsilon(4S)$ frame: $M_{\rm tag} = \sqrt{(E^2_{\rm
beam}-|{\mathbf p}_{\rm
tag}|^2})$ and 
$\Delta E_{\rm tag} =  E_{\rm tag} -E_{\rm beam}$, where $\hbox{\bf p}_{\rm tag} = \sum_i \hbox{\bf p}_i$, $E_{\rm tag} = \sum_i E_i$, $E_{\rm beam}$
is the colliding-beam energy, and ${\mathbf p}_i$ and $E_i$ denote the 
3-momentum
vector and
energy, respectively, of particle $i$.
The summation is over all particles that are
assigned to the $B_{\rm tag}$ candidate. We require that candidate events 
be in the
range $M_{\rm tag}>$ 5.2 GeV and
$-0.30$ ${\rm GeV}< \Delta E_{\rm tag}<0.05$ GeV.
The average number of candidates per event is about 1.03 for ($D^{\ast -}\ell^+$) pairs and 1.08 for ($D^{\ast -}\pi^+$) pairs. From multiple candidates, we select a ($D^{^\ast -}d^+_{\tau}$) pair
(throughout the paper, $d_{\tau}$  stands for the charged $\tau$ daughter:
$e$, $\mu$, or $\pi$) with the best $D^{\ast -}$ candidate, based on the
value of $\Delta M_{D^{\ast}}$. For the pairs sharing the same
$D^{\ast -}$
candidate, 
we perform a vertex fit to $B_{\rm tag}$ candidates,
using all charged particles assigned to the tagging side, and 
select the one with the largest fit probability. Events with incorrectly or incompletely reconstructed
$B_{\rm tag}$ are suppressed by imposing the following requirements: zero
total event charge; no charged leptons in $B_{\rm tag}$ decay; zero net
proton/antiproton number; $E_{\rm res}$, the residual energy in the
electromagnetic calorimeter ({\it i.e.}, the sum of energies of
clusters that are not
included in $B_{\rm sig}$ nor $B_{\rm tag}$) less than 0.8 GeV,
number of neutral particles on the tagging side $N_{\pi^0} +
N_{\gamma}<5$, and multiplicity of charged tracks
$N_{\rm ch}<15$.  For candidates
with $d_{\tau} = \pi$, we require no $K^0_L$ candidate  in the event. For further
analysis, we accept events that satisfy requirements derived from the
kinematics of signal decays:
$q^2\equiv M_W^2 = (p_{\rm sig}-p_{D^{\ast -}})^2 > 4$ ${\rm GeV}^2$,
($p_{\rm sig}= (E_{\rm beam}, -{\mathbf p}_{\rm tag})$),
and for
($D^{\ast -}\pi^+$) candidates, $M_W^2> M_{\rm miss}^2+m_{\tau}^2$
($m_{\tau}$ denotes the nominal mass of the $\tau$ lepton, and the
missing mass squared $M_{\rm miss}^2 = (p_{\rm sig} - p_{D^{\ast -}} -
p_{d_{\tau}^+})^2$ corresponds to the square of the effective
mass of the neutrino system).
With these requirements, the $M_{\rm tag}$ distribution of the 
signal peaks
at the $B^0$ mass with more than $80\%$ ($60\%$) of the events being
contained in the region $M_{\rm tag} > 5.26$ GeV for 
$d_\tau = \ell (\pi)$.
\section{\label{sec:backgr}Background suppression and calibration}
To suppress background, we exploit observables
that are sensitive to multiple neutrinos in the final state: the visible energy
$E_{\rm vis}$, which is the sum of the energies of all particles in the
event, and the fraction $X_{\rm miss} = (E_{\rm
miss}-|{\mathbf p}_{D^{\ast -}}+{\mathbf p}_{d_{\tau}^+}|)/
\sqrt{E_{\rm beam}^2-m_{B^0}^2}$, where $E_{\rm miss}=E_{\rm
beam}-(E_{D^{*-}}+E_{d_{\tau}^+})$ and $m_{B^0}$ is the nominal $B^0$
mass, that approximates missing mass and does not depend on the $B_{\rm tag}$ reconstruction \cite{AM}. 
$X_{\rm miss}$ falls in the range $[-1,1]$ for 
events with zero missing mass ({\it e.g.}, with a single neutrino) but takes larger values if there are more undetected particles
({\it e.g.}, multiple neutrinos).
The requirements $E_{\rm vis}<$ 8.7 (8.8)
GeV, and
$X_{\rm miss}>$ 1.5 (1.0) for $d_{\tau} = \ell$ ($\pi$)
are chosen to maximize the statistical figure of merit ${\rm FOM} = 
N_S/\sqrt{N_S+N_B}$
($N_S$ and $N_B$ denote the expected signal and background
yields in the window $M_{\rm tag} > 5.26$ GeV), assuming SM
value of $\mathcal{B}(B^0\to
D^{*-}\tau^+\nu_{\tau}) = 1.45\%$. 
The $M_{\rm tag}$
distributions are expected to be flat for most background components, 
while the
distribution of the signal remains unchanged. 
Residual peaking background
stems from semileptonic decays $B^0 \to D^{\ast -}\ell^+\nu_{\ell}$ and
$B \to D^{\ast -}X\ell^+\nu_{\ell}$ (including  $B \to
\bar{D}^{\ast\ast}\ell^+\nu_{\ell}$).   The abundance of  these and 
other background
components is  calibrated to data, 
separately
for each
signal decay chain.
The MC samples are divided into the following categories: $B \to
\bar{D}^{\ast}\ell^+\nu_{\ell}$,
$B\to \bar{D}^{\ast\ast}\ell^+\nu_{\ell}$, hadronic $B$ decays, and
$c\bar{c}$
and $u\bar{u} + d\bar{d} + s\bar{s}$ continuum.  Hadronic $B$ decays are
further split into two subcategories: events with correctly assigned
daughters to mother decays, and random combinatorial. 
The normalizations
of these components are determined 
by fitting the experimental
distributions
of $M_{\rm tag}$,  $\Delta E_{\rm tag}$,  $X_{\rm miss}$,
$E_{\rm miss}$, $E_{\rm vis}$, $E_{d_{\tau}^+}$, $M_W^2$, $E_{\rm res}$,
and
$R_2$, the last being the ratio of the second and zeroth Fox-Wolfram moments \cite{FW}.
The region $M_{\rm tag}>$ 5.26 GeV and 
$X_{\rm miss}> 0.75$ for leptonic $\tau$ decays, or
$X_{\rm miss}> 0.5$ for $\tau \to \pi \nu$, where we
expect enhanced signal contribution, is excluded from the fit.  In the
($D^{*-}\pi^+$) pairs, a large part of the background comes from fake 
$\bar{D}^0$ candidates. 
This component is fixed
from a comparison of the data and the MC data in the side-bands of 
the $m_{D^{0}}$  distributions.
Assuming the branching fraction $\mathcal{B}(B^0\to D^{\ast -}\tau^+\nu_{\tau}) = 
1.45\%$, 
we expect
in the signal enhanced region $M_{\rm tag} > 5.26$ GeV and
in the range $-1\leq \cos\theta_{\rm hel}\leq 0$
around 170 (200) signal (background) events for ($D^{\ast -}\ell^+$) pairs
and 115 (290) events for ($D^{\ast -}\pi^+$) pairs.\footnote{The region
$\cos\theta_{\rm hel}> 0$ is excluded from the analysis due to
strong
acceptance artifacts caused by the low $D^{\ast -}$  reconstruction
efficiency.}
In the latter case, the signal yield includes also cross-feed events from
other $\tau$ decays, mainly from $\tau^+ \to \rho^+ \bar{\nu}_{\tau}$.
For leptonic $\tau$ decays, the main background contribution ($\approx 35\%$) 
comes
from semileptonic $B$ decays to excited charmed resonances. 
Events with fake $\bar{D}^{(\ast)}$ candidates constitute around $45\%$ 
of the ($D^{\ast -}\pi^+$) background.
\section{Fitting procedure}
The $\cos\theta_{\rm hel}$ distribution is measured by dividing the range
$-1\leq \cos\theta_{\rm hel}\leq 0$ 
into three equidistant bins. Signal yield in the $I$-th bin of $\cos\theta_{\rm hel}$ is extracted from an extended unbinned maximum likelihood fit to the $M_{\rm tag}$ distributions in that bin 
using the following likelihood function:
\begin{equation}
\label{eq:likelihood}
\mathcal{L}^I = 
e^{-[N_s^I+\sum_k(N^k_{pI}+N^k_{bI})]}\prod_{i=1}^{N^I}[N_s^I\sum_k
w_k P_s^k(x_i)\\
+\sum_k (N_{pI}^k P_s^k(x_i)+N^k_{bI} P^k_{bI}(x_i))],
\end{equation}
where $x_i$ is the $M_{\rm tag}$ value of the $i^{\rm th}$ event,  
the index $k$ runs over decay chains, and
$N^I$ is the total number of events
in the $I^{\rm th}$ bin of $\cos\theta_{\rm hel}$
in  data. The probability density functions (PDF),
$P_s^k$, that describe signal and peaking background
are parameterized using the Crystal-Ball (CB) function \cite{TS}.
Our MC studies show that for a given type of $d_{\tau}$,
the shape of the CB component does not
depend on $\cos\theta_{\rm hel}$ and is the same for all $\bar{D}^0$
decays. PDFs denoted $P_{bI}^k$ describe
combinatorial background, and are parameterized
with an ARGUS-function (AR) \cite{Argus}.
All shape parameters of the PDFs are determined from fits to the MC
samples and fixed in the fit to data.
The coefficients
$w_k$ contain reconstruction efficiencies and partial decay rates of
individual decay chains, and are calculated using signal MC.  $N_s^I$,
$N_{bI}^k$, and $N_{pI}^k$ are the yields of signal, combinatorial
background
and peaking background in the $I^{\rm th}$ bin of $\cos\theta_{\rm hel}$,
respectively.
$N_s^I$ and $N_{bI}^k$ are
free parameters
of the fit, while $N_{pI}^k$ are fixed to the values obtained from fits to
MC samples and scaled to the data integrated luminosity.
The signal yields obtained in
the bins of $\cos\theta_{\rm hel}$ are
reweighted with the following scale factors $s_1 = 0.98\pm 0.01$
$(-1\leq\cos\theta_{\rm hel}<-0.67)$,
$s_2 = 0.96\pm 0.01$ $(-0.67\leq\cos\theta_{\rm hel}<-0.33)$,
and $s_3=1.08 \pm 0.01$ $(-0.33\leq\cos\theta_{\rm hel}<0)$,
in order to correct for small acceptance
variations along $\cos\theta_{\rm hel}$. 
The quoted errors arise from
statistical uncertainties of the signal MC.

The $D^{\ast -}$ polarization is measured by fitting the obtained
$\cos\theta_{\rm hel}$ distribution using Equation~\ref{eq:fl},
with $F_L^{D^\ast}$ as the only free parameter.
The procedure has been
tested by
fitting ensembles of simulated experiments varying
$F_L^{D^{\ast}}$
in the range of
$0\leq F_L^{D^{\ast}}\leq 1$. These pseudo-experiments are generated
using the shapes of the fitted PDFs for the signal and background
components and with the number of events Poisson-distributed around the
expected yields. The pull distributions of the extracted $F_L^{D^{\ast}}$
values are consistent with standard normal distributions in the entire
range of $F_L^{D^{\ast}}$.

As a cross check, we apply our procedure to measure the $D^\ast$
polarization in decays $B^0 \to D^{\ast-} e^+ \nu_e$, and obtain the
result $F_L^{D^\ast}(B^0 \to D^{\ast-} e^+ \nu_e)=0.56\pm 0.02$, which
agrees well with the value of 0.54 (0.53) 
predicted in the covariant quark 
model (heavy quark limit)
\cite{ivanov}.
\section{\label{sec:results}Results}

\begin{figure}[htb]

{
\includegraphics[width=1.0\textwidth]{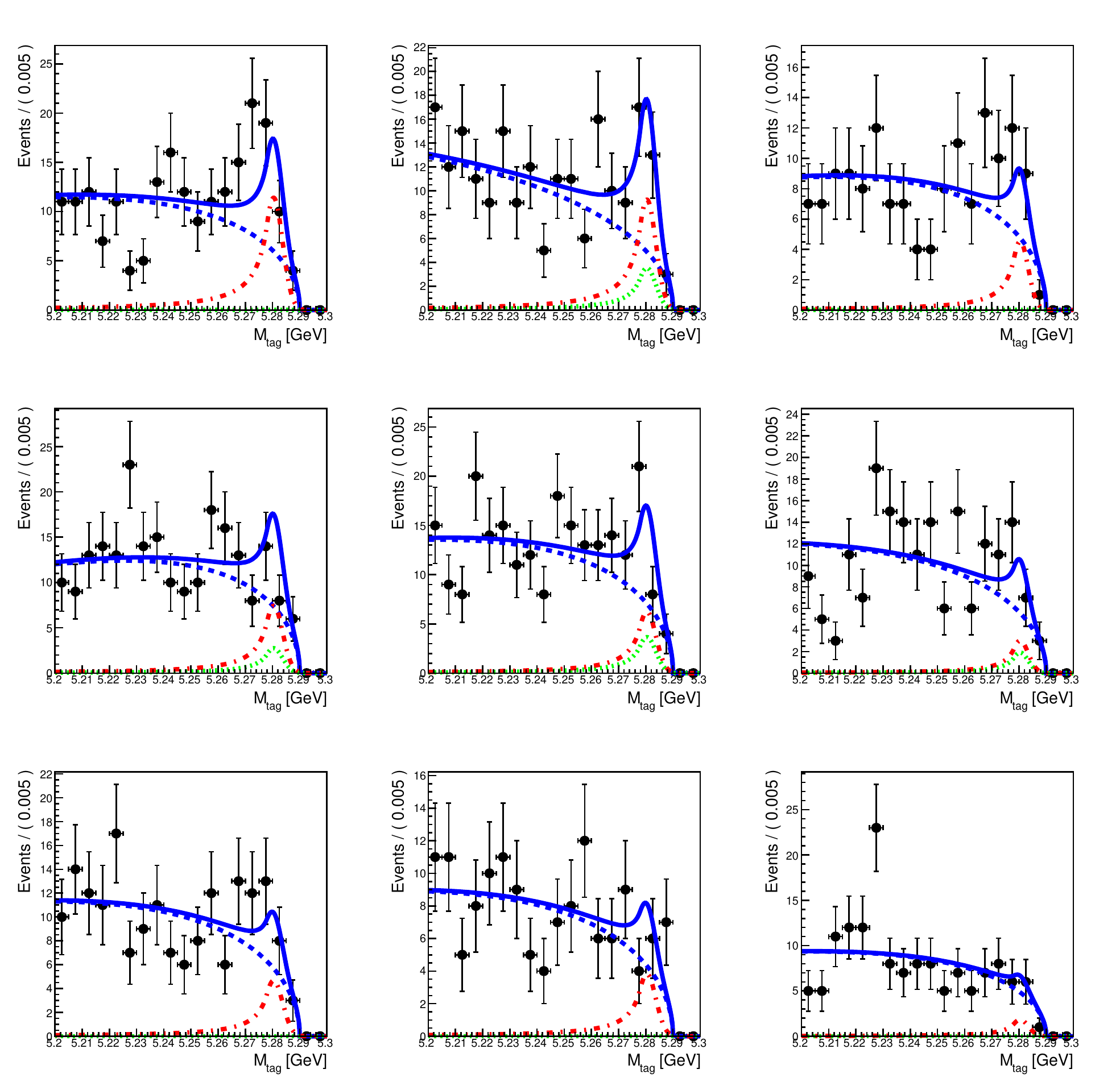}
}
\caption{Fit projections to $M_{\rm tag}$ distributions in three bins of $\cos\theta_{\rm hel}$ for $\tau \to \pi \nu_{\tau}$ (sequential columns) and $D\to K\pi$ (top), $D\to K\pi\pi^0$ (middle), $D\to K3\pi$ (bottom).
The solid lines show the result of the fit. 
Contributions of the signal, combinatorial and peaking backgrounds are
represented by the red (dot-dashed), blue (dashed) and green (dotted)
lines, respectively.
}
\label{fig:1}
\end{figure}

\begin{figure}[htb]

{
\includegraphics[width=1.0\textwidth]{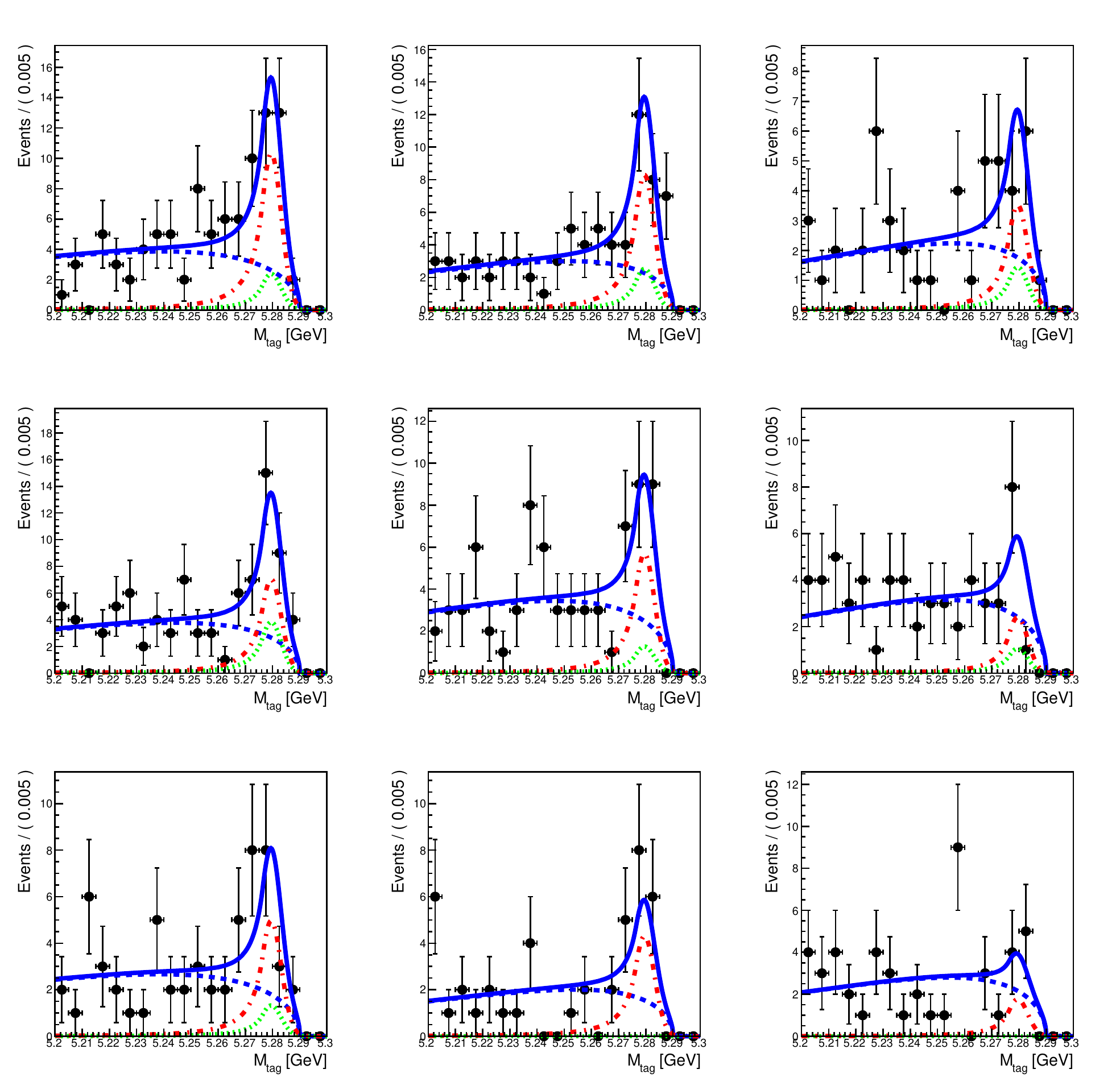}
}
\caption{Fit projections to $M_{\rm tag}$ distributions
in three bins of $\cos\theta_{\rm hel}$ for $\tau \to e {\bar\nu}_e \nu_{\tau}$ (sequential columns) and $D\to K\pi$ (top), $D\to K\pi\pi^0$ (middle), $D\to K3\pi$ (bottom). The solid lines show the result of the fit. 
Contributions of the signal, combinatorial and peaking backgrounds are
represented by the red (dot-dashed), blue (dashed) and green (dotted)
lines, respectively.
}
\label{fig:2}
\end{figure}

\begin{figure}[htb]

{
\includegraphics[width=1.0\textwidth]{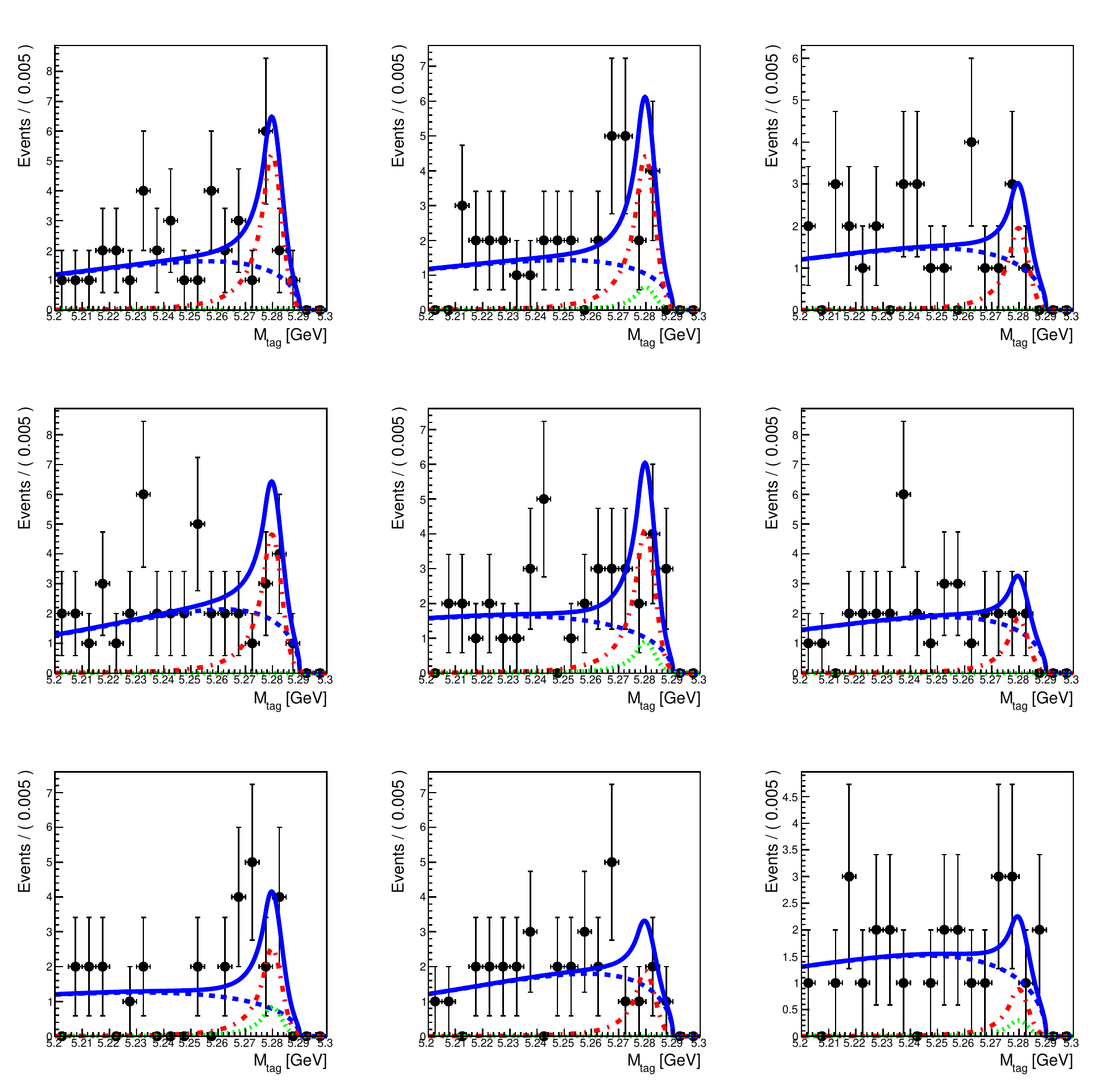}
}
\caption{Fit projections to $M_{\rm tag}$ distributions in three bins of $\cos\theta_{\rm hel}$ for $\tau \to \mu {\bar\nu}_{\mu} \nu_{\tau}$ (sequential columns) and $D\to K\pi$ (top), $D\to K\pi\pi^0$ (middle), $D\to K3\pi$ (bottom). The solid lines show the result of the fit. 
Contributions of the signal, combinatorial and peaking backgrounds are
represented by the red (dot-dashed), blue (dashed) and green (dotted)
lines respectively.
}
\label{fig:3}
\end{figure}

Applying the procedure described above to data, we obtain the following
yields of signal in the three bins of $\cos\theta_{\rm hel}$:
$N_s^1 = 151 \pm 21$ ($-1\leq\cos\theta_{\rm hel}<-0.67$),
$N_s^2 = 125 \pm 19$ ($-0.67\leq\cos\theta_{\rm hel}<-0.33$),
and $N_s^3 = 55 \pm 15$ ($-0.33\leq\cos\theta_{\rm hel}<0$), 
where the 
uncertainties are statistical.
The corresponding statistical significances are $\Sigma_1 =8.8 \sigma$, 
$\Sigma_2 
= 7.8 \sigma$, and $\Sigma_3 = 4.1 \sigma$, respectively.
The statistical significances are defined as
$\Sigma_I = \sqrt{-2\ln(\mathcal{L}_0^I/\mathcal{L}_{\rm max}^I)}$,
where $\mathcal{L}_{\rm max}^I$ and $\mathcal{L}_0^I$ denote the maximum
likelihood value and the likelihood values for the zero signal
hypothesis.
Fit projections are shown in Figs.~\ref{fig:1}--\ref{fig:3}. 

\begin{figure}[htb]

{
\includegraphics[width=0.45\textwidth]{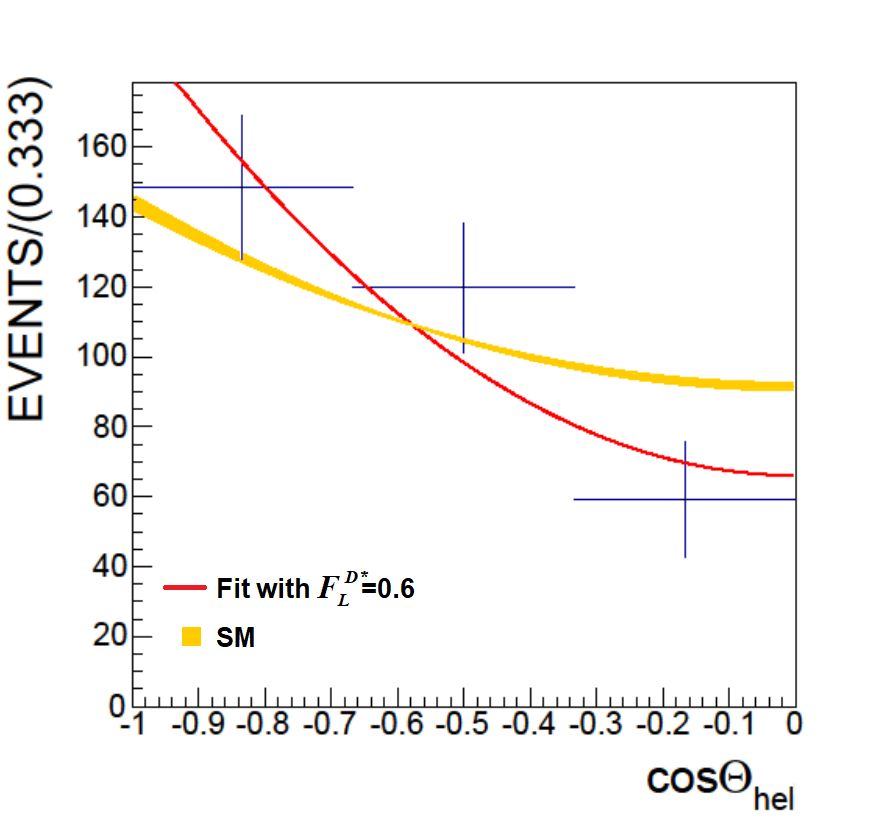}
}
\caption{The measured $\cos\theta_{\rm hel}$ distribution in  $B^0 \to
D^{\ast -} \tau^+\nu_{\tau}$ decays (data points with statistical errors); the fit 
result is overlaid (red line) with $F_L^{D^{\ast}} = 0.60$.
The yellow band represents the SM prediction of Ref.~\cite{huang}.
}
\label{fig:fl}
\end{figure}

The signal yields $N_s^I$ are weighted for acceptance corrections using
the scale factors $s_I$.
By fitting Eq.~\ref{eq:fl} to the obtained $\cos\theta_{\rm hel}$
distribution,
we measure $F_L^{D^{\ast}}=0.60 \pm 0.08$ (statistical) with $\chi^2/ndf = 1.95/2$.
The fit result is shown in Fig.~\ref{fig:fl}.
\subsection{\label{sec:level2}Systematic uncertainties}
The measurement of $F_L^{D^{\ast}}$ is not affected by absolute
normalization of the signal yield. Therefore, uncertainties related to the 
number of
$B\bar{B}$ pairs, $B_{\rm tag}$ reconstruction efficiency, and those
coming from the limited accuracy of the partial branching fractions
used in the analysis have no (or negligible) effect on the final
result.

The dominant systematic uncertainties arise from the
limited size of the MC sample and imperfect modelling of real 
processes.
They are summarized in Table~\ref{Tab:sys} and described below.

\begin{table}[htb]
\caption{Summary of systematic uncertainties}
\label{Tab:sys}
\begin{tabular}
{@{\hspace{0.5cm}}l@{\hspace{0.5cm}}|
@{\hspace{0.5cm}}l@{\hspace{0.5cm}}|@{\hspace{0.5cm}}c@{\hspace{0.5cm}}}
\hline \hline
Source & &$\Delta F_L^{D^{\ast}}$\\
\hline
Monte Carlo & AR shape and peaking background & $\pm 0.032$\\
statistics &CB shape &$\pm 0.010$ \\
 & Background scale factors& $ \pm 0.001$\\
\hline
Background & $B \to D^{\ast\ast}\ell\nu$ & $\pm 0.003$\\
modeling & $B \to D^{\ast\ast}\tau\nu$ & $\pm 0.011$\\
& $B \to {\rm hadrons}$ & $\pm 0.005$\\
& $B\to \bar{D}^\ast M$ & $\pm 0.004$\\
\hline
Signal modeling& Form factors & $\pm 0.002$\\
&$\cos\theta_{\rm hel}$ resolution & $\pm 0.003$\\
& Acceptance non-uniformity & $^{+0.015}_{-0.005}$\\

\hline
Total& & $^{+0.039}_{-0.037}$\\
\hline
\hline
\end{tabular}
\end{table}

To evaluate the effect of  statistical uncertainties of the
MC-determined
parameters that describe the shapes of the PDF, relative proportion of peaking
background, scale factors of background components, and correction
factors for the acceptance non-uniformities,
the procedure of $F_L^{D^{\ast}}$
measurement is repeated by varying each parameter 1000 times at
random,
assuming Gaussian errors, and taking into account correlations among them.
The standard deviation of the obtained $F_L^{D^\ast}$ distribution represents the
corresponding systematic uncertainty.

In the second category,
the poor knowledge
of semileptonic $B$ decays to excited charmed mesons,
representing a large part of the peaking background, is an
important source of systematic uncertainty.
For $B \to \bar{D}^{\ast\ast}\ell\nu$ modes, the branching fractions are varied for each $D^{\ast\ast}$ resonance
within
experimental uncertainties: $\pm 6\% (D_1)$, $\pm 10\%
(D_2^{\ast})$, $\pm 83\% (D_1')$, and $\pm 100\% (D^{\ast}_0,
D^{(\ast )}(2D))$ (assuming branching fractions of $0.5\%$ for modes
with radially excited states, $D^{(\ast )}(2S))$.
Uncertainties related to the form factor parameterization 
are negligible.
Uncertainty coming from  $B \to \bar{D}^{\ast\ast}\tau\nu$ decays, 
with the expected $\mathcal{B}(B \to 
\bar{D}^{\ast\ast}\tau\nu)\approx 0.3\%$,
is evaluated by changing their contribution by $\pm 100\%$ in the 
$B\bar{B}$
MC sample.

To estimate the uncertainties of combinatorial background from hadronic $B$ decays, we vary within $\pm 50\%$ the relative fractions
of 2-body, 3-body and n-body (n $ >3$) hadronic channels.
Two-body decays of the type $B \to \bar{D}^{\ast}M$,
where $M$ denotes a meson with a mass $M_M > 2 $ GeV,
with correctly assigned daughters to $B_{\rm sig}$ and $B_{\rm tag}$ 
decays, represent the main peaking background in the $\tau \to \pi
\nu_{\tau}$ mode. The systematic uncertainty
coming from the composition of the $M$ states (mainly the
$c\bar{s}$ resonances) is evaluated 
by reweighting
the $q^2$ spectrum by $\pm 100\%$ in two ranges of $q^2$: $q^2<6.2$ ${\rm GeV}^2$ and $q^2 > 6.2$ ${\rm GeV}^2$. (At $q^2 \approx 6.2$ GeV$^2$
there is a sharp change in the $\cos\theta_{\rm hel}$ distribution
for this component.)

Uncertainties due to the form factor
parameterization of signal decays are estimated by comparing
the results obtained with the two versions of the signal MC,
and found to be very small. 
Uncertainties related to the 
$\cos\theta_{\rm hel}$ resolution and acceptance non-uniformities along
$\cos\theta_{\rm hel}$ depend on the actual value of the $D^{\ast}$ 
polarization.
To evaluate them, the simulated signal events are reweighted 
to obtain $\cos\theta_{\rm hel}$ distributions that
correspond to arbitrary $D^{\ast}$ polarizations, and differences between the generated and measured values of $F_L^{D^\ast}$ are considered as systematic
uncertainties.
Uncertainties due to imperfect modeling of the $\cos\theta_{\rm hel}$ resolution
are within $\pm 0.003$ in the full range of $F_L^{D^\ast}$ values.
Variation of the $\cos\theta_{\rm hel}$ distribution affects the correction
factors $s_I$, resulting in the uncertainty
of $^{+0.015}_{-0.005}$
for the measured value of $F_L^{D^\ast} = 0.60$.


\section{\label{sec:conclusions}Conclusions}

We report the first measurement of the $D^\ast$ polarization in 
semitauonic
decay $B^0 \to D^{\ast -}\tau^+\nu_{\tau}$. The result is based on a data
sample of $772\times 10^6$
$B\bar{B}$ pairs collected with the Belle detector.
The fraction of $D^{\ast -}$ longitudinal polarization, measured assuming
SM dynamics, is found to be
$F_L^{D^\ast} = 0.60 \pm 0.08 ({\rm stat})\pm 0.04 ({\rm syst})$,
and agrees within 1.6 (1.8) standard deviations with the SM predicted 
values $(F_L^{D^\ast})_{\rm SM} = 
0.457 \pm 0.010$ \cite{Srimoy}
($0.441\pm 0.006$ \cite{huang}). 


\vspace{0.3cm}

We thank the KEKB group for the excellent operation of the
accelerator, the KEK cryogenics group for the efficient
operation of the solenoid, and the KEK computer group and
the National Institute of Informatics for valuable computing
and SINET3 network support. We acknowledge support from
the Ministry of Education, Culture, Sports, Science, and
Technology of Japan and the Japan Society for the Promotion
of Science; the Australian Research Council and the
Australian Department of Education, Science and Training;
the National Natural Science Foundation of China under
contract No.~10575109 and 10775142; the Department of
Science and Technology of India; 
the BK21 program of the Ministry of Education of Korea, 
the CHEP SRC program and Basic Research program 
(grant No.~R01-2005-000-10089-0) of the Korea Science and
Engineering Foundation, and the Pure Basic Research Group 
program of the Korea Research Foundation; 
the Polish State Committee for Scientific Research; 
the Ministry of Education and Science of the Russian
Federation and the Russian Federal Agency for Atomic Energy;
the Slovenian Research Agency;  the Swiss
National Science Foundation; the National Science Council
and the Ministry of Education of Taiwan; and the U.S.\
Department of Energy.


\begin{thebibliography}{99}
\bibitem{CG} Throughout this paper, the inclusion of the charge-conjugate
mode decay is always implied.
\bibitem{units} Natural units system with $\hslash = c= 1$ is used.
\end{thebibliography}
\end{document}